\documentclass{article}

\usepackage{PRIMEarxiv}

\usepackage[utf8]{inputenc} 
\usepackage[T1]{fontenc}    
\usepackage{hyperref}       
\usepackage{url}            
\usepackage{booktabs}       
\usepackage{amsfonts}       
\usepackage{nicefrac}       
\usepackage{microtype}      
\usepackage{lipsum}
\usepackage{fancyhdr}       
\usepackage{graphicx}       
\graphicspath{{media/}}     

\title{\textbf{Evidence of Galactic Center Expansion from Gaia DR3 Stars along the South–North Axis: Implications for the JWST Early Galaxy Problem} 
\textbf{
}
}
\author{G.S. Karapetian\textsuperscript{1*},
L.E. Byzalov\textsuperscript{3}, M. A. Hovhannisyan\textsuperscript{2}, L.A. Mahtessian\textsuperscript{2}
and
A.P. Mahtessian\textsuperscript{1}
 \\
  Affiliation \\
\textsuperscript{1} Byurakan Astrophysical Observatory after V. Ambartsumian NAS of the Republic \\ of  Armenia,  Byurakan, Aragatzotn Province , Republic of Armenia, 0213  \\
\textsuperscript{2 }Institute of Applied Problems of Physics NAS of the Republic of Armenia
25 Hrachya Nersissian Str.,\\ Yerevan, Republic of Armenia, 0014\\
\textsuperscript{3 } University of Waterloo, Ontario, Canada\\  200 University Ave W, Waterloo, ON N2L 3G1\\
  \texttt{* garenk53@gmail.com} \\
}

\begin{document}
\maketitle

\begin{abstract}
Recent observations with the James Webb Space Telescope (JWST) have revealed massive, evolved galaxies only a few hundred million years after the Big Bang, challenging standard cosmological models. To test the hypothesis that the Galactic nucleus may act as a source of matter and energy capable of accelerating galaxy formation, we examined whether the Milky Way itself shows signs of central activity. Expansion of the Galactic center would imply such activity in its nuclear region. To investigate this possibility, we analyzed stellar motions along the south–north axis of the Galaxy using Gaia DR3 data.

Average Galactocentric radial velocities were computed in 50 control points, spaced by 0.1 kpc with a sampling radius of 0.05 kpc, up to 5 kpc from the Galactic center. The results show a strong symmetry: in the northern region, 36 of 50 points have positive velocities, with an average of +19.15 ± 10.80 km/s (N = 50), while in the southern region, 37 of 50 points have negative velocities, with an average of –19.24 ± 8.22 km/s (N = 50). A Student’s t-test confirms that the two distributions differ significantly ($p \approx 0.003$).

 The combination of positive average velocities in the north and negative average velocities in the south indicates a symmetric outflow of stars away from the Galactic center, consistent with expansion in its central region.
 
These Galactocentric radial velocities are consistent with the formation of massive, evolved galaxies within $\sim 300$ My, as observed by JWST. If extended to other galaxies, the results — together with earlier findings on globular cluster motions — suggest that aspects of standard galaxy formation models may require refinement. This interpretation, however, remains preliminary and requires further study.
\end{abstract}


\section{Introduction}
With the release of the Gaia DR3  \cite{Gaia2022DR3}, it has become possible to test hypotheses and ideas that could not previously be examined due to the limited size of earlier stellar catalogs of our Galaxy. One such hypothesis is the assumption that the Milky Way is undergoing expansion. There are relatively few studies in the literature devoted to this topic. For example,  \cite{Mart´ınez2019} reported evidence of disk expansion in galaxies similar to the Milky Way, with measured velocities of several hundred meters per second. However, the authors attribute this effect to disk truncation and stellar migration, noting that it occurs only at the outer edges of galactic disks and is not directly related to the central regions.

The objective of this study is to investigate the potential expansion of the central region of the Milky Way by analyzing stellar motions with respect to the Galactic center. In our previous work  \cite{Karapetian2024GC}, we used globular clusters, representing a population of the Galaxy’s second type, as the objects of study. It was shown that, on average, these clusters are moving away both from the Galactic plane and from the center, with average recession velocities ranging from 17 to 31 km/s.

\section{Method}
\subsection{Determination of control points (CP) of the study.}
As noted above, the study was conducted for stars located within 0.05 kpc of the south–north axis of the Galaxy. This approach was chosen because, in this case, the calculation of Galactocentric radial velocities does not require the $radial\_velocity$ parameter for stars, as is the case when using the $v\_radial\_gc$ function of the Astropy package \cite{Astropy2018}. This makes it possible to include a significantly larger number of objects in the analysis, since in Gaia DR3 only a relatively small fraction of stars have measured radial velocities. In addition, this selection method simplifies the correction for the oscillation velocity of the Sun’s orbit in the direction of the Galactocentric z coordinate.

For the analysis, control points (CPs) were chosen along this axis with a step of 0.1 kpc. Around each CP, a spatial region with a radius of about 0.05 kpc was defined (see Fig. 1). Within each such region, the average Galactocentric radial velocities of the stars were calculated.

\begin{figure}[hbt!]
    \centering
    \includegraphics[width=0.8\linewidth]{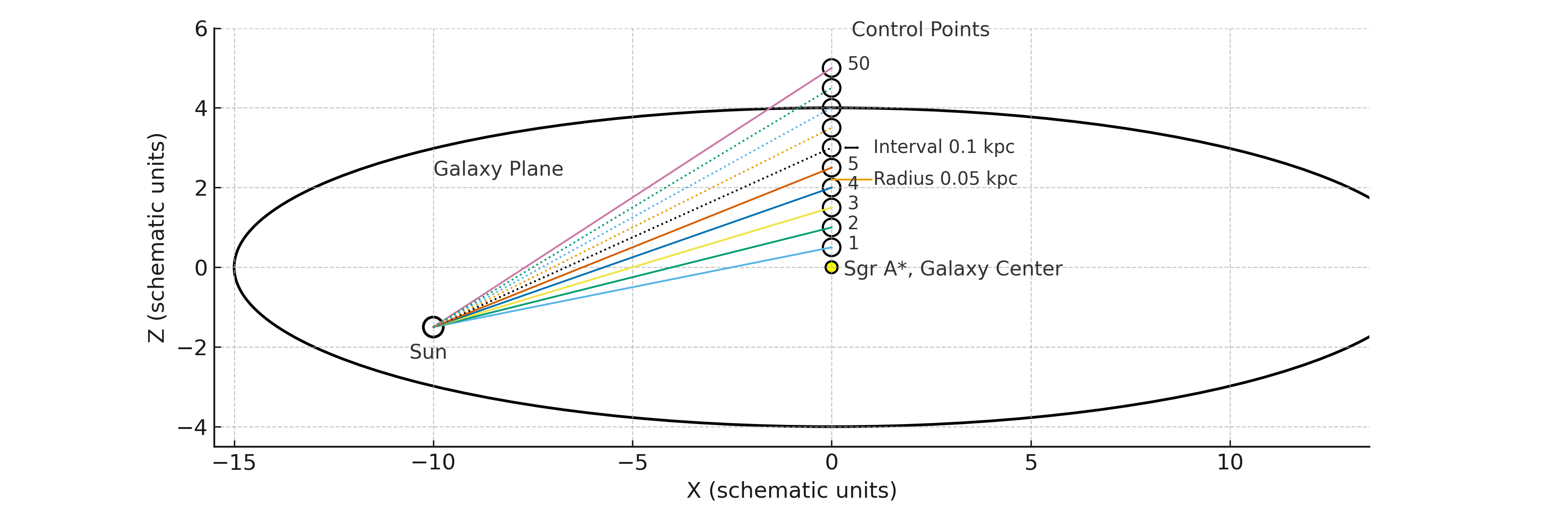}
    \caption{Schematic layout of control points (CPs) relative to the Galaxy.}
  \end{figure}

If the average velocity values in the control points (CPs) are close to zero, this can be interpreted as the absence of expansion. Conversely, positive values in the northern region would indicate a possible expansion of both the Galactic halo and, potentially, the central part of the Galaxy. Negative values in the northern region, on the other hand, may be interpreted as evidence of contraction within distances up to 5 kpc from the center. The same criteria are applicable to the southern region of the Galaxy.

Symmetric control points (CPs) were also defined in the southern region of the Galaxy. For each CP, stellar parameters were retrieved from the Gaia DR3 archive (https://gea.esac.esa.int/archive/) using appropriate filters, corresponding to the volumes of space surrounding each point.

The search in the Gaia DR3 archive gaiadr3.gaia\_source was performed using the following CP parameters:

\hspace*{7mm}• $\alpha$   - the right ascension of the center of the search area (CP);

\hspace*{7mm}• $\delta$ - the declination of the center of the search area (CP);

\hspace*{7mm}• SR - Search radius (the radius of the search cone in arcsec);

\hspace*{7mm}• CP - Close Parallax (parallax of the nearest surface of the CP region);

\hspace*{7mm}• FP – Far Parallax (far surface parallax of the CP region).

These parameters defined a region with a radius of approximately 0.05 kpc around the center of each CP, from which stellar data were retrieved. The calculation of CP parameters is presented in Table 1 in the Appendix.

Since the majority of stars from the Gaia DR3 catalog located along the south–north axis are concentrated within 5 kpc of the Galactic center, the analysis was carried out for 50 CPs. The sample size of stars in individual CPs varied from several hundred near the center to several at distances greater than 3 kpc. In total, 6670 stars were selected for the southern region and 3433 stars for the northern region. The dependence of the number of stars in the CPs on the distance from the Galactic center for the northern region is shown in Fig. 2.

\begin{figure}
    \centering
    \includegraphics[width=0.6 \linewidth]{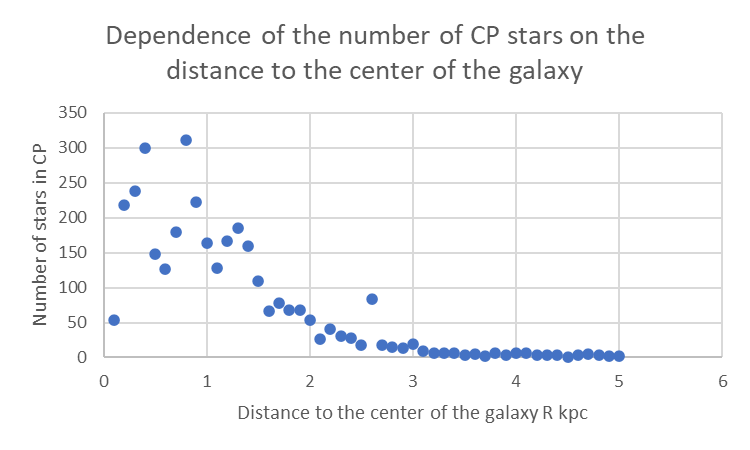}
    \caption{The number of northern CP stars depending on the distance to the center of the Galaxy }
    \label{fig:enter-label}
\end{figure}

The following star parameters were downloaded from the archive:

\hspace*{7mm}• designation

\hspace*{7mm}• ra ($\alpha$)

\hspace*{7mm}• dec ($\delta$)

\hspace*{7mm}• Parallax

\hspace*{7mm}• pm (Proper motion)

\hspace*{7mm}• pmra ($\mu$$\alpha$) – Proper motion in the direction of right ascension

\hspace*{7mm}• pmdec ($\mu$$\delta$) – Proper motion in the direction of declination

\hspace*{7mm}• radial\_velocity (RV)

\hspace*{7mm}• phot\_g\_mean\_mag – G – band mean magnitude

\hspace*{7mm}• phot\_bp\_mean\_mag – Integrated BP mean Magnitude

\hspace*{7mm}• phot\_rp\_mean\_mag – Integrated RP mean Magnitude 

\subsection{Calculation of Galactocentric Radial Velocities for CP Stars.}
The Galactocentric radial velocity of a star, $V_{\mathrm{gc}}$, was determined as follows. The distance of the star to the Galactic center was computed for two epochs --- at the time of observation (Gaia DR3) and one year later. The difference between these distances, divided by the time interval, provided an estimate of the star’s velocity relative to the Galactic center.

Although Galactic coordinates $l$ and $b$ are provided in the Gaia database, for consistency and improved accuracy we recalculated them independently using the appropriate coordinate transformations. Thus, for each star in a control point, both the present coordinates $(l, b)$ and the coordinates one year later $(l', b')$ were determined.

Based on these values, the Galactocentric radial velocity was calculated over one year (in kpc\,yr$^{-1}$) using the following formula:

\begin{equation}
V_{gs} = R_{\odot} \left[ \tan(b + \mu_b) - \tan(b) \right] 
      = R_{\odot} \left[ \tan(b') - \tan(b) \right]
\end{equation}

where:  
\begin{itemize}
    \item $V_{gs}$ is the velocity of the star relative to the center of the Galaxy in kpc\,yr$^{-1}$,  
    \item $R_{\odot}$ is the distance from the Sun to the Galactic center.  
\end{itemize}

 When averaging the velocities, we also accounted for the vertical oscillation of the Sun’s orbit relative to the Galactic plane. The adopted value of this velocity was $W_{\odot}$=9.3 km/s, following the estimate of  \cite{Gordon2023} which refines the result obtained by \cite{Reid2020}, \cite{Xu2022}, \cite{Francis2022}, \cite{Dehnen1998} and \cite{Schonrich2010}.

Fig. 3 and Fig. 4 show the dependence of the average Galactocentric velocities 
$V^{\uparrow}_{cp,i}$ and $V^{\downarrow}_{cp,i}$ on the distance to the Galactic center.

As can be seen, the average velocities of most CPs in the northern region have positive values (36 out of 50) and indicate, on average, the receding of stars from the center of the Galaxy in the vertical direction relative to the plane in the interval from the center 0-5 kps with an average velocity  of +19.15±10.80 km/s (n=50), while for the southern region they have negative values (37 out of 50) and are moving away from the center of the Galaxy with an average speed of -19.24±8.22 km/s (n=50).

\begin{figure}[hbt!]
    \centering
    \includegraphics[width=0.6\linewidth]{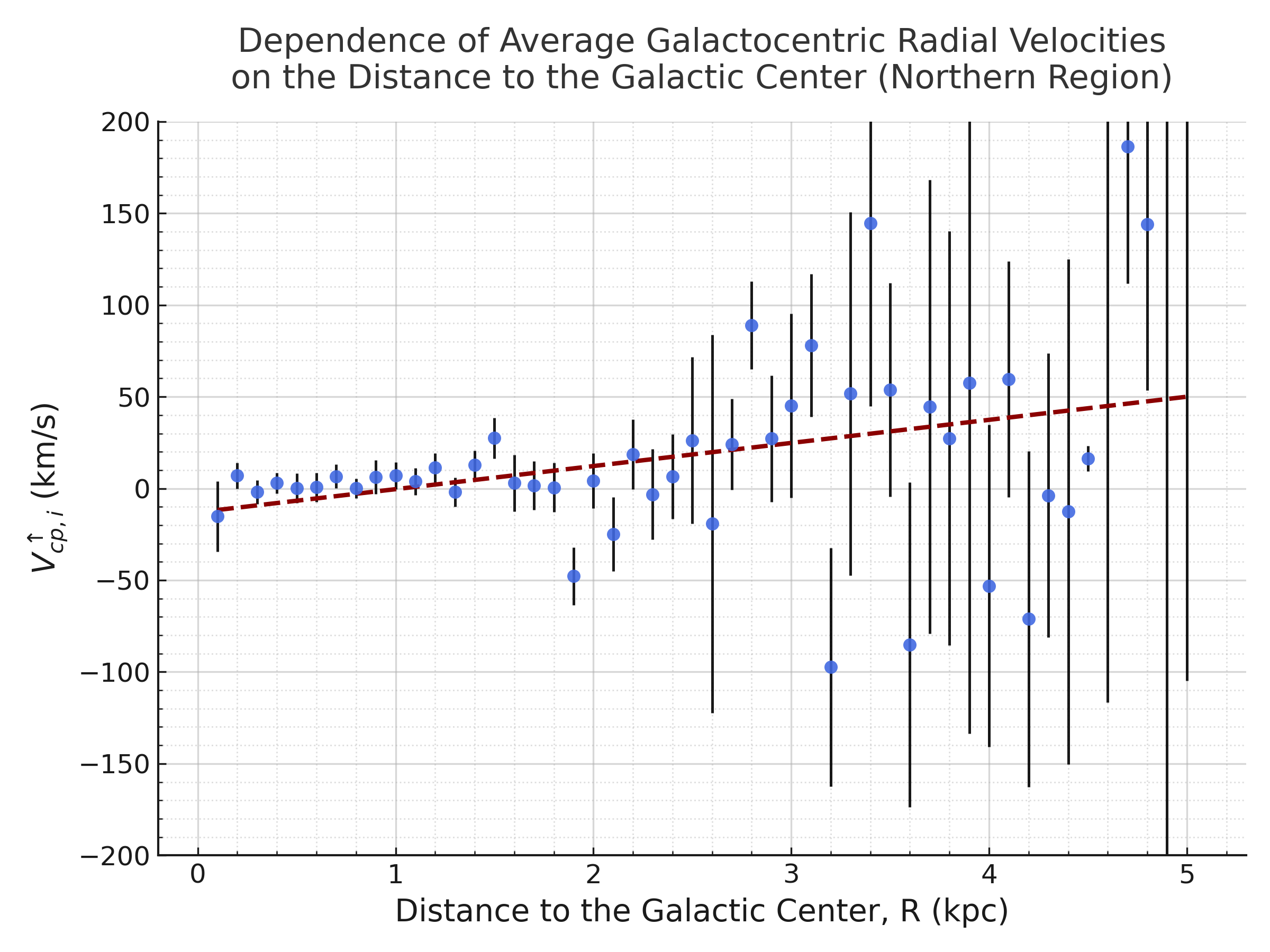}
    \caption{The average velocities of most CPs in the northern region have positive values (36 out of 50) and, accordingly, indicate movement away from the center of the Galaxy with an average velocity of 19.15±10.80   km/s.
 }
   \end{figure}
   
\begin{figure}[hbt!]
    \centering
    \includegraphics[width=0.6\linewidth]{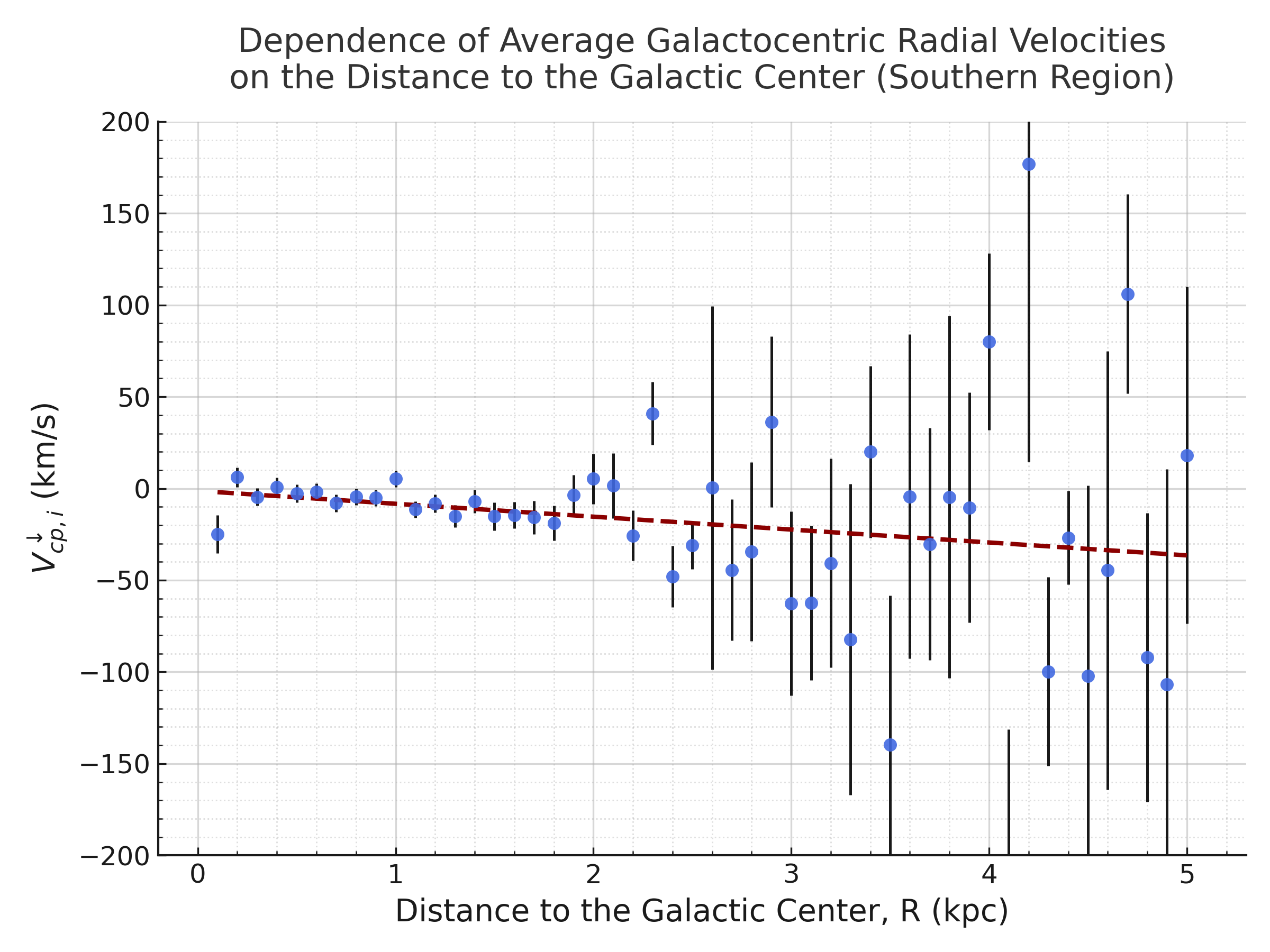}
    \caption{The average velocities of most control points (37 out of 50) in the southern region are negative, indicating motion away from the Galactic center with a mean velocity of -19.24±8.22   km/s. 
 }
   \end{figure}

Applying Student’s t-test showed that the distributions of  $V^{\uparrow}_{cp,i}$ and $V^{\downarrow}_{cp,i}$  for the northern and southern regions differ statistically at a significance level of $ p \approx0.003$.

The revealed symmetry of the velocities $V^{\uparrow}_{cp,i}$ and $V^{\downarrow}_{cp,i}$ is consistent with the observed spatial symmetry of the Galactic halo along the south–north axis. In the case of a significant asymmetry in velocities, this symmetry would be disrupted on cosmological timescales.

\section{Results}

We obtained the following averaged values of the Galactocentric radial velocities of stars in the control points (CPs):

In the northern region:

\begin{itemize}
    \item 0–5 kpc: +19.15±10.80 km/s (N=50)
    \item 1–5 kpc: +23.19±13.14 km/s (N=41)
    \item 2–5 kpc: +30.10±17.15 km/s (N=31)
    \item 3–5 kpc: +37.38±24.92 km/s (N=21)
    \item 4–5 kpc: +42.24±44.02 km/s (N=11)
\end{itemize}

In the southern region:

\begin{itemize}
    \item $0$--$5$ kpc: $-19.24 \pm 8.22\,\mathrm{km/s}$ (N=50)
    \item $1$--$5$ kpc: $-22.37 \pm 9.96\, \mathrm{km/s}$ (N=41)
    \item $2$--$5$ kpc: $-26.21 \pm 13.12\, \mathrm{km/s}$ (N=31)
    \item $3$--$5$ kpc: $-33.97 \pm 18.72\, \mathrm{km/s}$ (N=21)
    \item $4$--$5$ kpc: $-26.89 \pm 33.87\, \mathrm{km/s}$ (N=11)
\end{itemize}

\section{Conclusions \& Discussion}
The results of this study indicate that the Milky Way exhibits expansion along the south–north axis within $\sim 5$ kpc of the Galactic center. This is consistent with \cite{Karapetian2024GC}, who found that globular clusters are receding from both the Galactic plane and the center with average velocities of 17–31.7 km/s. Together, these findings suggest that expansion may occur not only in the halo but also in the central region of the Galaxy. It would also be useful to analyze Population I and II stars separately; however, this task is complicated by the lack of data in the Gaia DR3 archive that would allow for reliable correction of the observed BP–G and G–RP color indices for interstellar extinction. 

Thus, the widely held assumption that the Galaxy is not expanding is not supported here. This does not necessarily imply an overall growth of Galactic size, as the expansion may diminish with radius and could even turn into contraction at larger distances.

The apparent expansion of the central region, pointing to the presence of an active nucleus capable of accelerating galaxy formation, remains in tension with standard theory and should be regarded as preliminary and requires further study. 

Nevertheless, such an interpretation may help explain JWST detections of massive galaxies younger than 1 Gyr (\cite{Xiao2024},\cite{Labbé2023}, \cite{Naidu2022}, \cite{Finkelstein2023}). At an expansion rate of $\sim 20$ km s$^{-1}$, a galaxy could reach a diameter of $\sim 10$ kpc within 300 Myr, consistent with the sizes of early massive galaxies observed by JWST.

Further research may be conducted by incorporating a larger number of stars from different quadrants and segments of the Galaxy.

\bibliographystyle{unsrt}  
\bibliography{references}  

\appendix
\section{Appendix}
\begin{table}
\caption{Parameters of Control Points of the Northern Region of the galaxy}
\centering
\begin{tabular}{l c c c c c c c c l}
CP & \textit{b}

\textit{(deg)} & R (Kpc) & Dist.

to CP 

 & Par-x 

(MAS) &$\alpha$

(deg) & $\delta$ 

(deg) & SR

(AS) & FP

(MAS) & CP

(MAS) \\
\hline
1 & \textit{0.72} & 0.1 & 7.99763 & 0.12504 & 265.7087712 & -28.56110 & 1290 & 0.12426 & 0.12581 \\
2 & \textit{1.43} & 0.2 & 7.99950 & 0.12501 & 265.0177301 & -28.18262 & 1289 & 0.12423 & 0.12578 \\
3 & \textit{2.15} & 0.3 & 8.00263 & 0.12496 & 264.3320128 & -27.80094 & 1289 & 0.12418 & 0.12573 \\
4 & \textit{2.86} & 0.4 & 8.00700 & 0.12489 & 263.6517521 & -27.41625 & 1288 & 0.12412 & 0.12567 \\
5 & \textit{3.58} & 0.5 & 8.01262 & 0.12480 & 262.9770724 & -27.02878 & 1287 & 0.12403 & 0.12558 \\
6 & \textit{4.29} & 0.6 & 8.01948 & 0.12470 & 262.3080903 & -26.63872 & 1286 & 0.12392 & 0.12547 \\
7 & \textit{5.00} & 0.7 & 8.02758 & 0.12457 & 261.6449139 & -26.24628 & 1285 & 0.12380 & 0.12534 \\
8 & \textit{5.71} & 0.8 & 8.03692 & 0.12443 & 260.9876433 & -25.85169 & 1283 & 0.12366 & 0.12520 \\
9 & \textit{6.42} & 0.9 & 8.04748 & 0.12426 & 260.3363707 & -25.45514 & 1282 & 0.12350 & 0.12503 \\
10 & \textit{7.13} & 1 & 8.05928 & 0.12408 & 259.6911802 & -25.05685 & 1280 & 0.12332 & 0.12485 \\
11 & \textit{7.83} & 1.1 & 8.07230 & 0.12388 & 259.0521485 & -24.65703 & 1278 & 0.12312 & 0.12464 \\
12 & \textit{8.53} & 1.2 & 8.08653 & 0.12366 & 258.4193443 & -24.25588 & 1275 & 0.12290 & 0.12442 \\
13 & \textit{9.23} & 1.3 & 8.10198 & 0.12343 & 257.7928293 & -23.85362 & 1273 & 0.12267 & 0.12418 \\
14 & \textit{9.93} & 1.4 & 8.11862 & 0.12317 & 257.1726577 & -23.45044 & 1270 & 0.12242 & 0.12393 \\
15 & \textit{10.62} & 1.5 & 8.13646 & 0.12290 & 256.5588768 & -23.04655 & 1268 & 0.12215 & 0.12365 \\
16 & \textit{11.31} & 1.6 & 8.15549 & 0.12262 & 255.9515269 & -22.64215 & 1265 & 0.12187 & 0.12336 \\
17 & \textit{12.00} & 1.7 & 8.17570 & 0.12231 & 255.350642 & -22.23743 & 1261 & 0.12157 & 0.12306 \\
18 & \textit{12.68} & 1.8 & 8.19707 & 0.12199 & 254.7562493 & -21.83259 & 1258 & 0.12126 & 0.12273 \\
19 & \textit{13.37} & 1.9 & 8.21961 & 0.12166 & 254.1683701 & -21.42781 & 1255 & 0.12092 & 0.12240 \\
20 & \textit{14.04} & 2 & 8.24330 & 0.12131 & 253.5870197 & -21.02328 & 1251 & 0.12058 & 0.12204 \\
21 & \textit{14.71} & 2.1 & 8.26813 & 0.12095 & 253.0122075 & -20.61917 & 1247 & 0.12022 & 0.12167 \\
22 & \textit{15.38} & 2.2 & 8.29409 & 0.12057 & 252.4439377 & -20.21567 & 1243 & 0.11985 & 0.12129 \\
23 & \textit{16.05} & 2.3 & 8.32118 & 0.12018 & 251.8822092 & -19.81293 & 1239 & 0.11946 & 0.12089 \\
24 & \textit{16.71} & 2.4 & 8.34937 & 0.11977 & 251.3270156 & -19.41114 & 1235 & 0.11906 & 0.12048 \\
25 & \textit{17.36} & 2.5 & 8.37866 & 0.11935 & 250.7783462 & -19.01044 & 1231 & 0.11864 & 0.12006 \\
26 & \textit{18.01} & 2.6 & 8.40904 & 0.11892 & 250.2361854 & -18.61099 & 1226 & 0.11822 & 0.11962 \\
27 & \textit{18.66} & 2.7 & 8.44050 & 0.11848 & 249.7005134 & -18.21294 & 1222 & 0.11778 & 0.11917 \\
28 & \textit{19.30} & 2.8 & 8.47302 & 0.11802 & 249.1713065 & -17.81643 & 1217 & 0.11733 & 0.11871 \\
29 & \textit{19.93} & 2.9 & 8.50659 & 0.11756 & 248.648537 & -17.42160 & 1212 & 0.11687 & 0.11824 \\
30 & \textit{20.56} & 3 & 8.54119 & 0.11708 & 248.1321736 & -17.02858 & 1207 & 0.11640 & 0.11776 \\
31 & \textit{21.19} & 3.1 & 8.57683 & 0.11659 & 247.6221815 & -16.63750 & 1202 & 0.11592 & 0.11727 \\
32 & \textit{21.81} & 3.2 & 8.61348 & 0.11610 & 247.118523 & -16.24847 & 1197 & 0.11543 & 0.11677 \\
33 & \textit{22.42} & 3.3 & 8.65113 & 0.11559 & 246.6211571 & -15.86162 & 1192 & 0.11493 & 0.11626 \\
34 & \textit{23.03} & 3.4 & 8.68976 & 0.11508 & 246.1300403 & -15.47705 & 1187 & 0.11442 & 0.11574 \\
35 & \textit{23.64} & 3.5 & 8.72938 & 0.11456 & 245.6451263 & -15.09486 & 1181 & 0.11390 & 0.11521 \\
36 & \textit{24.24} & 3.6 & 8.76995 & 0.11403 & 245.1663664 & -14.71515 & 1176 & 0.11338 & 0.11467 \\
37 & \textit{24.83} & 3.7 & 8.81147 & 0.11349 & 244.6937099 & -14.33801 & 1170 & 0.11285 & 0.11413 \\
38 & \textit{25.42} & 3.8 & 8.85393 & 0.11294 & 244.227104 & -13.96353 & 1165 & 0.11231 & 0.11358 \\
39 & \textit{26.00} & 3.9 & 8.89730 & 0.11239 & 243.7664937 & -13.59179 & 1159 & 0.11177 & 0.11302 \\
40 & \textit{26.57} & 4 & 8.94159 & 0.11184 & 243.3118229 & -13.22287 & 1153 & 0.11122 & 0.11246 \\
41 & \textit{27.14} & 4.1 & 8.98677 & 0.11127 & 242.8630334 & -12.85682 & 1148 & 0.11066 & 0.11189 \\
42 & \textit{27.71} & 4.2 & 9.03283 & 0.11071 & 242.4200659 & -12.49373 & 1142 & 0.11010 & 0.11132 \\
43 & \textit{28.27} & 4.3 & 9.07976 & 0.11014 & 241.9828598 & -12.13364 & 1136 & 0.10953 & 0.11074 \\
44 & \textit{28.82} & 4.4 & 9.12754 & 0.10956 & 241.5513533 & -11.77662 & 1130 & 0.10896 & 0.11016 \\
45 & \textit{29.37} & 4.5 & 9.17617 & 0.10898 & 241.1254837 & -11.42271 & 1124 & 0.10839 & 0.10957 \\
46 & \textit{29.91} & 4.6 & 9.22562 & 0.10839 & 240.7051874 & -11.07196 & 1118 & 0.10781 & 0.10898 \\
47 & \textit{30.44} & 4.7 & 9.27588 & 0.10781 & 240.2904 & -10.72441 & 1112 & 0.10723 & 0.10838 \\
48 & \textit{30.97} & 4.8 & 9.32695 & 0.10722 & 239.8810567 & -10.38010 & 1106 & 0.10664 & 0.10779 \\
49 & \textit{31.50} & 4.9 & 9.37881 & 0.10662 & 239.4770918 & -10.03906 & 1100 & 0.10606 & 0.10719 \\
50 & \textit{32.02} & 5 & 9.43144 & 0.10603 & 239.0784395 & -9.70131 & 1093 & 0.10547 & 0.10659 \\

\end{tabular}

\end{table}

\end{document}